\documentclass[final,5p,times,twocolumn]{elsarticle}

\usepackage[english]{babel} 				
\usepackage[utf8]{inputenc}					
\usepackage[T1]{fontenc}    				
\usepackage{textgreek}    				

\usepackage{amssymb}
       
\journal{Journal of Alloys and Compounds}

\begin{document}

\begin{frontmatter}



\title{Electronic and magnetic ground state of MnB$_{4}$}


\author[tubs]{N.Steinki\corref{cor1}\fnref{fn1}}
\ead{n.steinki@tu-bs.de}
\author[tubs]{J.L. Winter}
\author[tubs]{D. Schulze Grachtrup}
\author[tubs]{D. Menzel}
\author[tubs]{S. Süllow}
\author[tudm]{A. Knappschneider}
\author[tudm]{and B. Albert}

\address[tubs]{Institut für Physik der Kondensierten Materie, TU Braunschweig, 38106 Braunschweig, Germany}
\address[tudm]{Eduard-Zintl-Institut für Anorganische und Physikalische Chemie, TU Darmstadt, 64287 Darmstadt, Germany}

\begin{abstract}
Recent studies have dealt with the electronic and magnetic ground state properties of the tetraboride material MnB$_4$. So far, however, the ground state properties could not be established unambiguously. Therefore, here we present an experimental study on single-crystalline MnB$_4$ by means of resistivity and magnetization measurements. For this, we have developed a sample holder that allows four-point ac resistivity measurements on these very small ($\sim$\,100\,$\mu$m) samples. With our data we establish that the electronic ground state of MnB$_4$ is intrinsically that of a pseudo-gap system, in agreement with recent band structure calculations. Furthermore, we demonstrate that the material does neither show magnetic order nor a behavior arising from the vicinity to a magnetically ordered state, this way disproving previous claims. 
\end{abstract}


\end{frontmatter}


\section{INTRODUCTION}
\label{introduction}

For decades now, the chemistry and physics of boron compounds has attracted the interest of a large community of researchers \cite{gmelin,albert}. Essentially, in compounds the element boron has a tendency towards network formation, which in turn translates into a complex interplay of structural, electronic and magnetic properties in these materials. Examples range from superhard materials such as FeB$_4$ \cite{gou2013} via superconductors with very high transition temperatures (MgB$_2$) \cite{nagamatsu} or unusual ferromagnets and antiferroquadrupolar systems (EuB$_6$, CeB$_6$) \cite{sullow,lee} to highly unusual surface states in the Kondo insulator SmB$_6$ \cite{zhang}.

A case in point is MnB$_4$: While being known to exist since 1960 \cite{fruchart}, recently the boride system MnB$_4$ became the focus of more extensive research efforts \cite{wang,gou2012,litterscheid,yang,knappschneider1,gou2014,niu,geest,liang,knappschneider2,LiangWuWang}. Initially, scientific interest was triggered by the search for superhard materials \cite{wang,gou2012,yang,knappschneider2,LiangWuWang}. However, as result of the first studies it appeared that the structural properties of MnB$_4$ are closely connected to its electronic properties, leading to the performance of a number of calculations (mostly by density functional theory) on the interdependency of electronic and crystallographic structure \cite{knappschneider1,gou2014,niu,geest,liang,knappschneider2,LiangWuWang}.

Surprisingly, with respect to the experimental verification of the different scenarios concerning the electronic structure of MnB$_4$, there is a marked inconsistency reported in literature. In Ref.~\cite{gou2014} in a thermodynamic study on polycrystalline material Gou et al. reported that MnB$_4$ exhibits ferromagnetic correlations (based on susceptibility) and a metallic ground state (based on specific heat). In contrast, Knappschneider et al. \cite{knappschneider1} investigated single crystals and concluded that the material - because of a Peierls-type distortion - should be considered a non-magnetic small-gap semiconductor or pseudo-gap material. From these experimental studies it appears that poly- and single crystalline material behave differently, bringing up the issue of sample quality and the intrinsic magnetic and electronic ground state of MnB$_4$. Based on recent band structure calculations \cite{liang,LiangWuWang} it has been attempted to put these issues into the context of a complex interplay of - as the authors call it - Peierls and Stoner mechanisms.

\begin{figure}
\centering
		\includegraphics[width=0.4\textwidth]{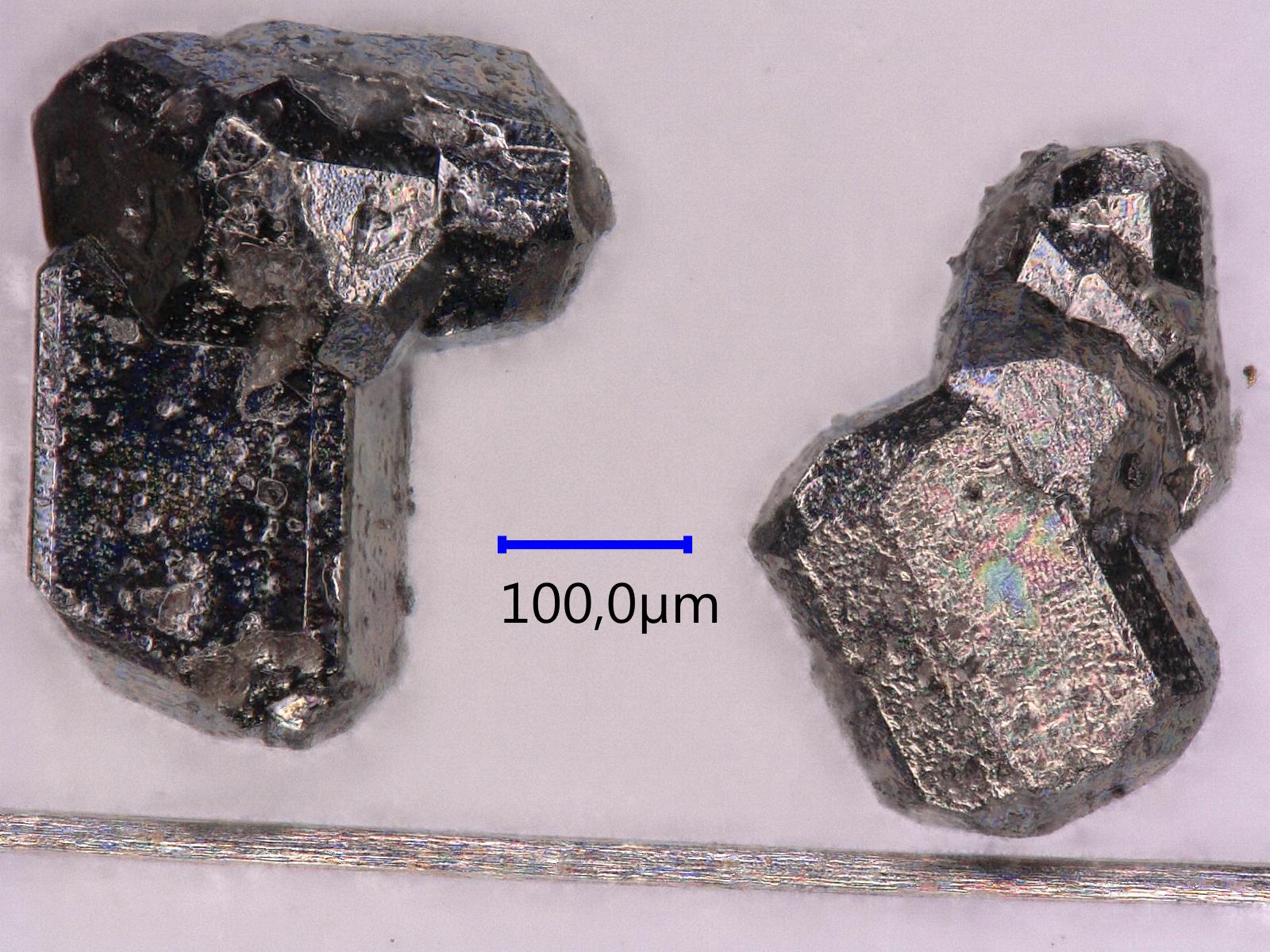}
	\caption{Two single-crystalline specimens of MnB$_4$ produced as described in Ref.~\cite{knappschneider1} as representatives of the samples studied in this work. For size comparison, one of the Pt wires used for the resistivity measurements is shown at the bottom of the figure.}
	\label{sample}
\end{figure}

Experimentally, the issue at hand is the availability of only very small single-crystalline specimens MnB$_4$ obtained from the synthesis. This is exemplified in Fig.~\ref{sample}, where we plot typical single-crystalline samples MnB$_4$ studied in the present work \cite{crystal}. These crystals have been produced according to the recipe described in Ref.~\cite{knappschneider1}. 

The limitations regarding the available samples has hampered experimental investigations in a way that a very basic characterization such as a four-probe ac resistivity measurement has not been reported. Only, in Ref.~\cite{knappschneider1} preliminary two-probe experiments on single-crystals MnB$_4$ in a limited temperature range $\sim 300 - 360$\,K have been presented and taken as evidence of a small semiconducting gap $\sim 0.04$\,eV in the density of states.

Correspondingly, what is lacking is a four-probe ac resistivity characterization of single-crystalline MnB$_4$ in a wider temperature range and down to low temperatures, in order to verify the notion of either a metallic or non-metallic ground state of the material. As well, such a study would allow insight into the issue of impurities affecting the physical properties. Therefore, we have set out performing this investigation by developing an experimental stage to carry out a temperature dependent four-probe resistive measurement on crystals of the order of down to $\sim 100$\,$\mu$m length, and using this tool to characterize the electronic ground state of MnB$_4$. Furthermore, we have accompanied our study by measuring the magnetization of single-crystalline MnB$_4$ in order to test the proposal of a ground state close to ferromagnetism.

\section{EXPERIMENTAL}
\label{experimental}

The common approach to measure the resistivity of a bulk sample is to use a four-point probe configuration. For this, a mechanical and electrical connection between sample and measurement wires is necessary, and which is commonly provided by use of different types of bonding, by fixing the wires with silver epoxy or paint onto the sample surface, or by mechanically pressing contacts pads onto the sample. 

Attachment of the contacts on the sample surface by bonding or soldering depends on the composition and morphology of the sample surface. For arbitrary and arbitrarily shaped samples and for typical bonding stations, although in principle small contacts can be produced this way, it will usually not result in stable contacts. Conversely, the option of gluing the wires to the sample surface is limited by the size of the sample because the silver epoxy/paint contacts typically have a diameter of the order of part of a mm. In this situation it is not possible to attach four wires to samples of a similar size. Finally, in recent years microfabricated contact pads have become commercially available allowing four-point resistivity measurements on small samples via mechanical contacts, but only over limited temperature ranges \cite{microres,capres}. 

\begin{figure}
    \centering
		\includegraphics[width=0.3\textwidth]{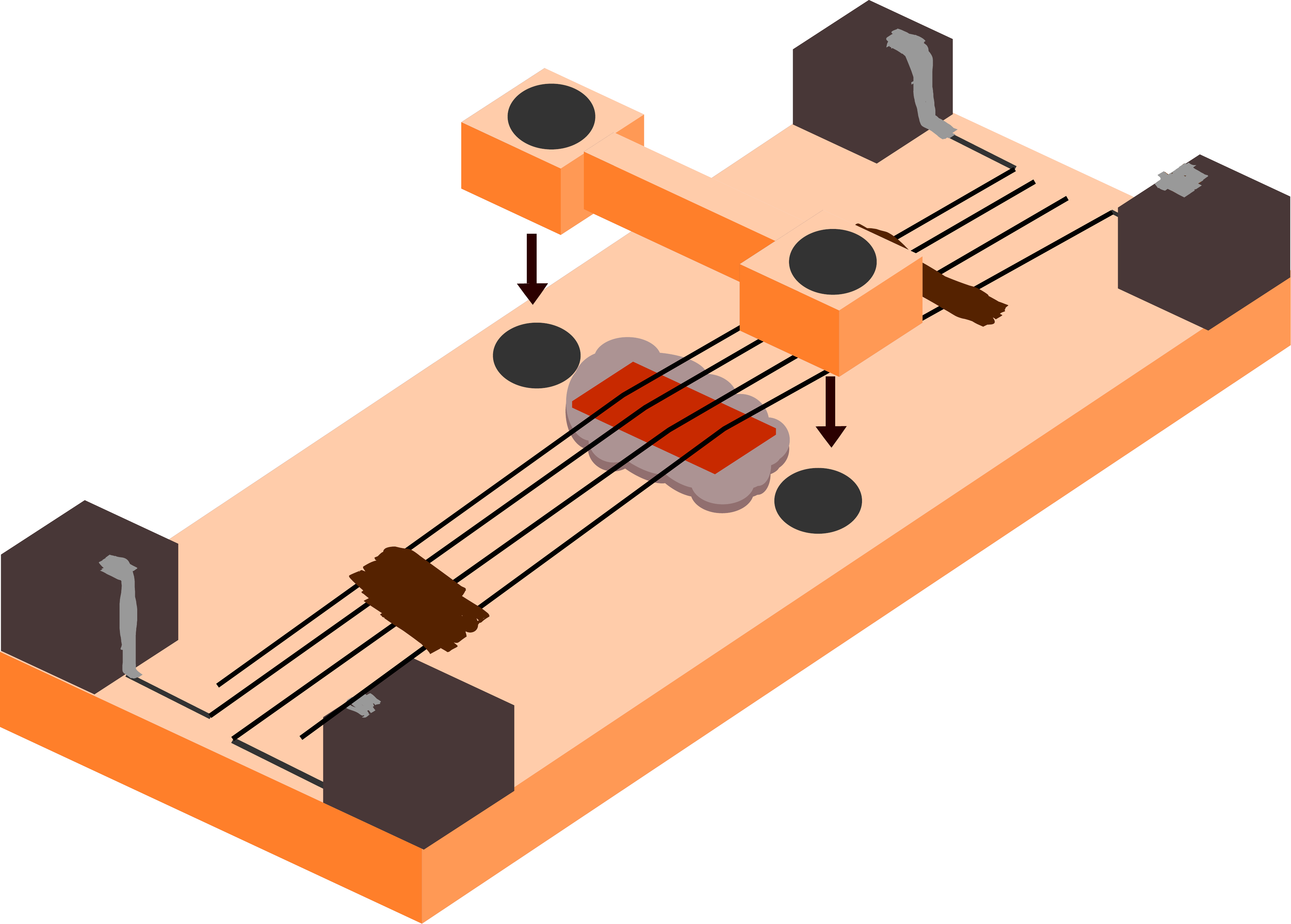}
		\includegraphics[trim=10cm 5cm 0 0, clip, width=0.3\textwidth]{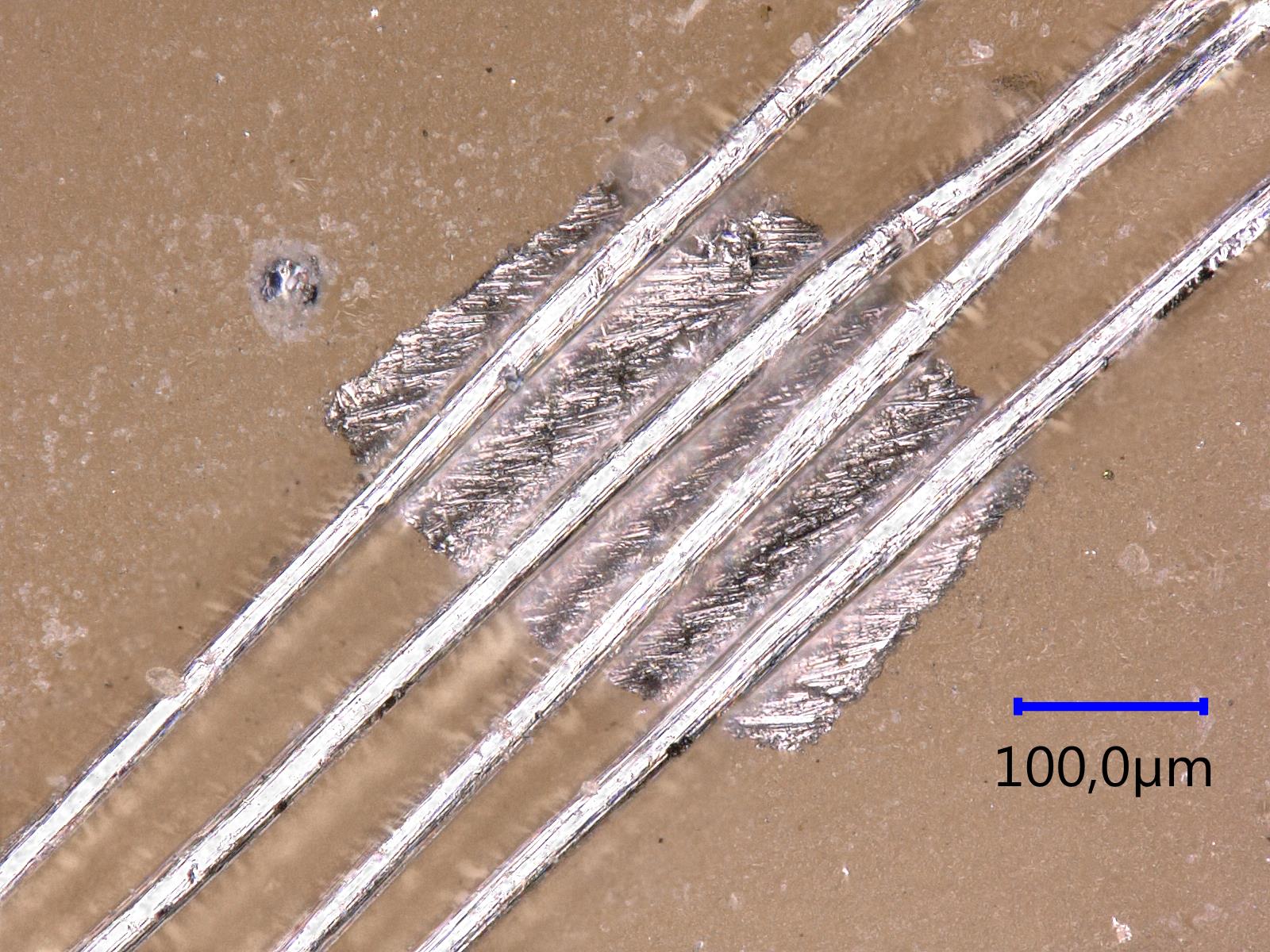}
	\caption{Schematic drawing of the sample holder for four-point probe resistivity measurements on small samples (top), together with a photograph of a sample (a piece of vanadium) prepared for a measurement using the holder (bottom).}
	\label{sampleholder}
\end{figure}

In this situation we have developed a simple and cheap sample holder for temperature dependent resistivity measurements, essentially based on the experimental knowledge from high pressure studies \cite{eremets}. Our sample holder allows attachment to standard laboratory set-ups for four-point probe resistivity measurements on crystals down to a length of about 100 $\mu$m and above (see Fig.~\ref{sampleholder}).

For the mechanical handling of the small samples, these are embedded in electrically insulating epoxy, allowing positioning in defined locations. For providing electrical contact, one side of the samples is exposed by sanding the epoxy/sample block (see bottom of Fig.~\ref{sampleholder} for a view of the resulting sample surface, in this case a piece of vanadium). After fixing the epoxy/sample unit on an insulated copper sample holder, four 25 $\mu$m platinum wires are drawn over the free sample surface in a guitar string fashion (top of Fig. \ref{sampleholder}). To produce electrical contact between wires and sample an electrically insulating piston presses the wires onto the sample surface. This way, it is also guaranteed that the wires do not move. We have tested the functionality of the set-up through measurements of various metallic samples (Cu, V, Nb). Altogether, with this setup it is possible to use the common four-point probe resistive configuration for samples with a length down to - in our case - 150\,$\mu$m, and for our cryogenic systems at temperatures below $\sim$\,200\,K.

Next, the sample holder is attached to a standard measurement stick for experiments in a $^4$He bath cryostat, with the platinum wires connected to the wiring for the resistivity measurements. Subsequently, we have determined the resistivity of various single-crystalline samples MnB$_4$ in a temperature range between 4.2 and 200\,K. With the embedding of the sample in epoxy, we have no control of the alignment of the crystals, and thus can not measure the resistivity along a specific crystallographic direction. In addition, magnetization measurements on the crystals have been carried out using a commercial SQUID magnetometer.           

With the irregular shape of the MnB$_4$ single-crystals, and taking into account that the sanding removes a small, but not well-defined part of the sample, in terms of resistivity measurements there is some uncertainty regarding the sample cross section. As well, the width of the voltage contacts (25\,$\mu m$) effectively defines the experimental error of the determination of the distance between these contacts, as they represent a significant portion of the contact distance (see Fig.~\ref{sampleholder}). Both geometrical factors have a quite large impact on the accuracy of the determination of the absolute value of the specific resistivity with this set-up. For the experiments presented here we have estimated that there is a 25\% error margin for the absolute values of the resistivity. The relative accuracy of the measurement is essentially determined by the electronics and easily below 1\%.

\section{RESULTS AND DISCUSSION}
\label{results}

\begin{figure}
\centering
		\includegraphics[width=0.45\textwidth]{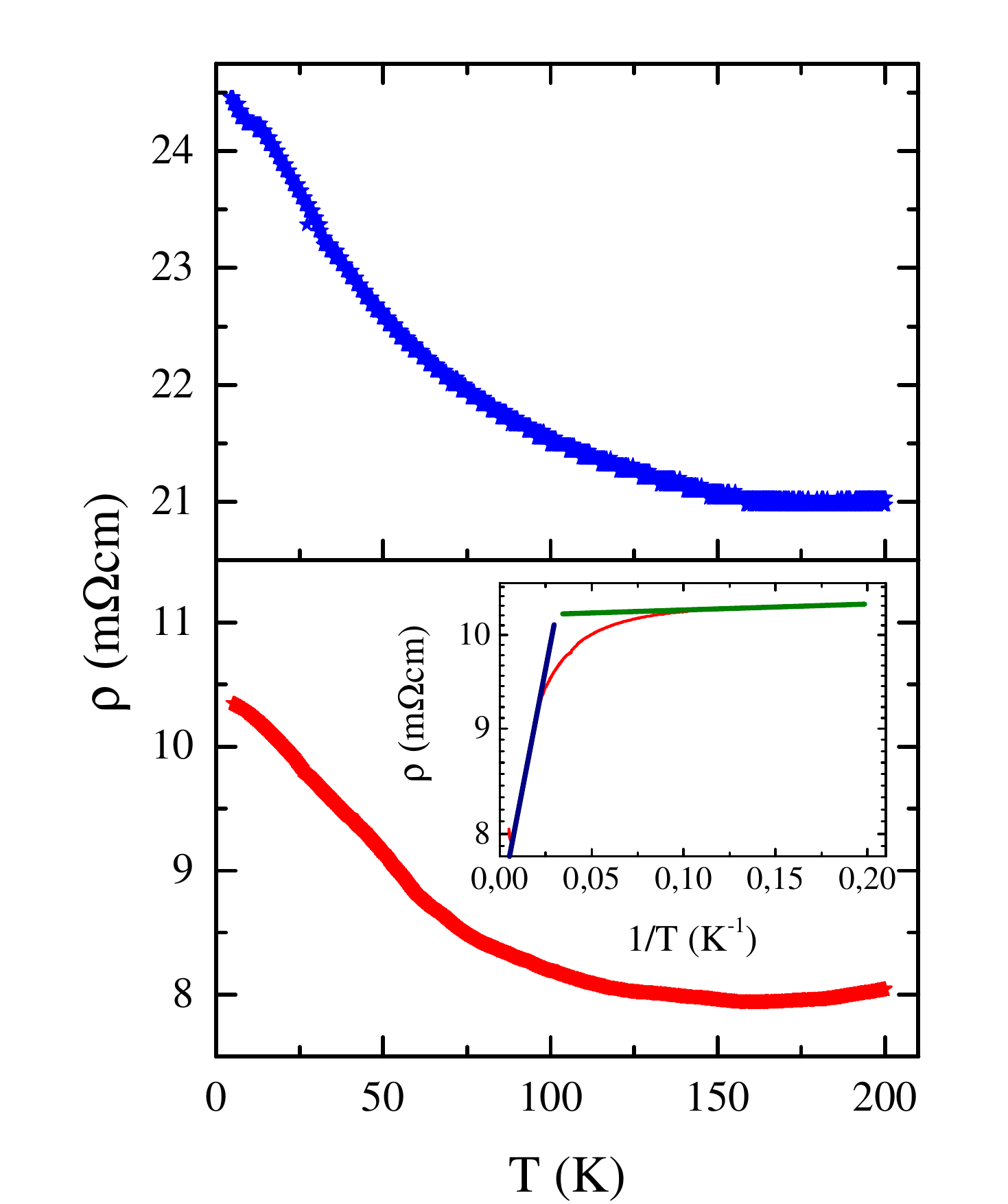}
	\caption{The resistivity of two single-crystalline samples MnB$_4$; the inset illustrates the attempts to derive an energy gap value from the data, for details see text.}
	\label{herte}
\end{figure}

In Fig.~\ref{herte} we present the resistivity for two single-crystalline samples MnB$_4$. Qualitatively, both exhibit a similar behavior, with overall a temperature dependence inconsistent with a metallic character for MnB$_4$. In detail, there are some differences between the measurements. First and most notably, the absolute values of the resistivity of the two samples differ by a factor 2.5, and which is beyond the margin of experimental error as described above. It thus reflects a sample dependence of the resistivity.

Given the variation in absolute resistivity values between different samples, comparing our data with the two-point probe data from Knappschneider et al.~\cite{knappschneider1}, we observe that also the absolute resistivity values from these previous data are broadly consistent with our study. In other words, we find the resistivity (conductivity) of single-crystalline MnB$_4$ to be of the order of a few ten m$\Omega$cm (a few hundred $S$m$^{-1}$), and rising (falling) by about 20\% as temperature is lowered into the $^4$He-range. At temperatures above 150\,K the resistivity is flat, and for one sample actually slightly increases with temperature.

The experimentally observed resistivity does not reflect archetypical semiconducting behavior. Yet, if - for the sake of the argument - we assume that the resistivity does reflect semiconductivity, it would imply that using an Arrhenius plot we can derive the corresponding gap values. Therefore, in the inset of Fig.~\ref{herte} we include the Arrhenius plot for the sample with the smaller ($\sim 10$\,m$\Omega$cm) resistivity. Clearly, there is no extended temperature range were the data can be approximated by a straight line. We can try to estimate upper and lower limits for energy gaps by linearly approximating the experimental data at lowest and highest measured temperature (see straight lines in the inset of Fig.~\ref{herte}). This approach would yield gap values between a few $10^{-6}$ to $10^{-4}$\,eV, which clearly is far too small to make sense.

Altogether, while the resistivity of single-crystalline MnB$_4$ does not exhibit a straightforward metallic character, it can also not be accounted for in terms of a simple semiconductor. Distinct from semiconductors, the absolute resistivity is much smaller (order of ten m$\Omega$cm rather than many $\Omega$m for typical semiconductors), the temperature dependence is much weaker (less than one order of magnitude change of $\rho$ with two orders of magnitude variation of temperature), and at high temperatures even tends towards a positive, {\it i.e.}, metallic resistive slope $d \rho /d T$.

The most likely explanation for this behavior is fully consistent with the predictions from band structure calculations as set out in Knappschneider et al. \cite{knappschneider1}. According to these DFT calculations the Peierls-type distortion induces a pseudo-gap in the density of states at the Fermi level of MnB$_4$. The presence of a pseudo-gap itself will give rise to a semiconducting-like resistivity, {\it i.e.}, the resistivity increases with decreasing temperature. The observation of a pronounced sample dependence points towards the presence of impurities with energy levels in the range of the pseudo-gap. Combined, pseudo-gap and impurities will produce a resistive behavior intermediate between metallicity and semiconductivity (for comparison see for instance the case of Fe$_{3-x}$V$_x$Al, and how impurities affect the transport properties in such a pseudo-gap sample series \cite{nishino,matsushita,maksimov}). For related borides the possibility of ''self-doping'' effects has been long established \cite{albert,amsler,aronson}.

\begin{figure}
\centering
		\includegraphics[width=0.45\textwidth]{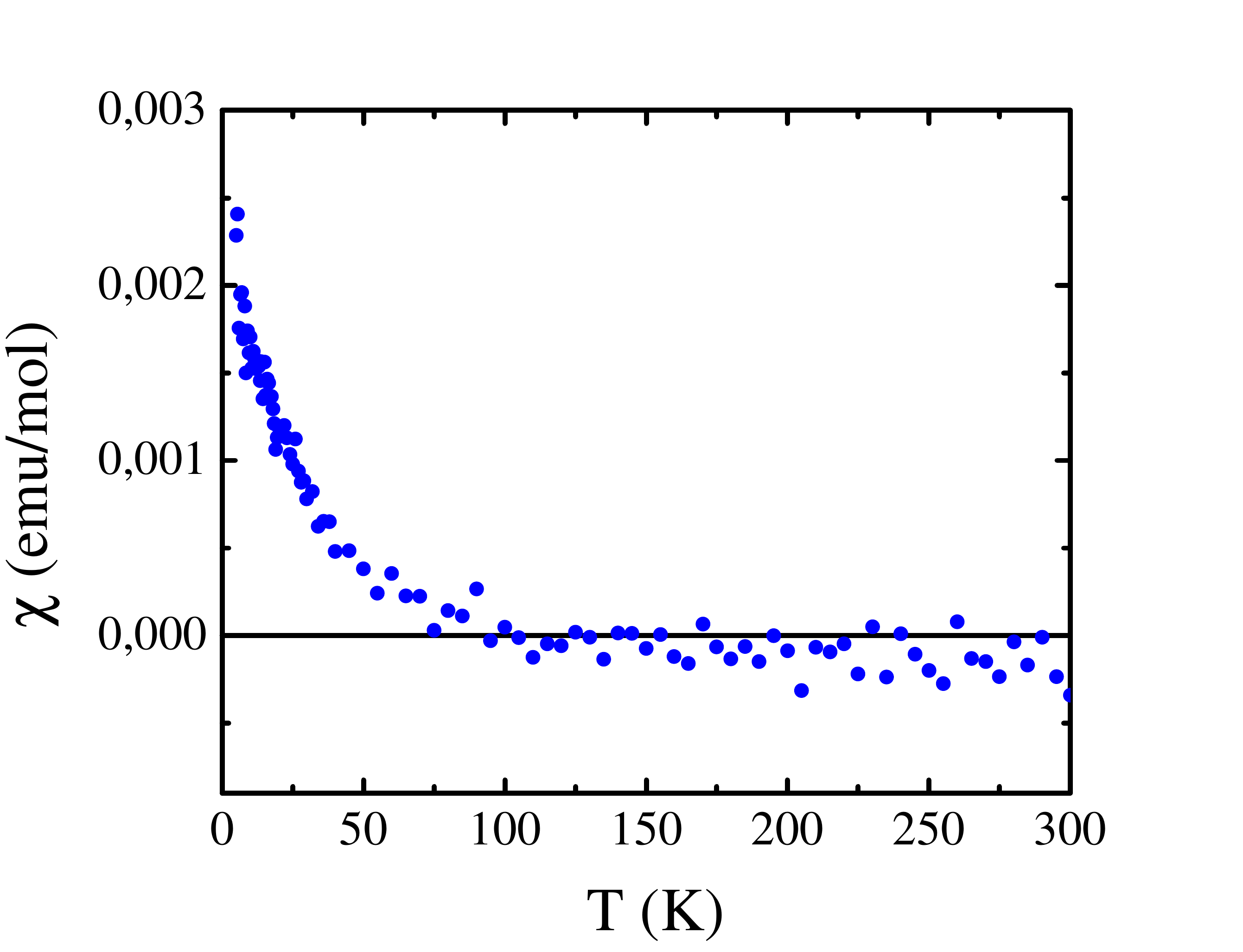}
	\caption{The temperature dependence of the susceptibility of single crystalline MnB$_4$, for details see text.}
	\label{susceptibility}
\end{figure}

Within this scenario of impurity states in a pseudo-gap in the band structure of MnB$_4$, the question arises about the extrinsic or intrinsic nature of the ferromagnetic signatures reported in Ref.~\cite{gou2014}. Therefore, in Fig.~\ref{susceptibility} we plot the temperature dependence of the susceptibility of single-crystalline MnB$_4$ (weight of the sample: 30 $\mu g$, applied field $B = 0.1$\,T). In the data analysis, we have subtracted the diamagnetic contribution of the Apiezon grease used to attach the sample to the straw from the raw data. Above 100\,K the measured signal is very small, {\it i.e.}, of the order of the resolution limit of the SQUID. This finding is at variance with the observations made for the polycrystalline material studied in Ref.~\cite{gou2014}. At low temperatures, the susceptibility becomes paramagnetic below 100\,K, although the absolute value of the susceptibility is still more than one order of magnitude smaller than in Ref.~\cite{gou2014}.

As a first approach to interpret the susceptibility of single-crystalline MnB$_4$, we start by assuming that the magnetic signal is fully intrinsic to the material. This would imply that the very small signal at high temperatures ($> 100$\,K) reflects an intrinsic property of MnB$_4$ (possibly diamagnetism), which would be consistent with a pseudo-gap in the band structure, but inconsistent with Gou et al. \cite{gou2014}. In turn, still assuming an intrinsic nature of the susceptibility, the change to a paramagnetic behavior at low temperatures would then correspond to a phase transition into a ferromagnetic state, but with an extremely small magnetic moment of the order of about 1/1000th of a $\mu_B$ per Mn atom. Both the concept of a ferromagnetic state appearing out of diamagnetism as well as the small moment size make this interpretation unlikely. This scenario leads to contradictions, which can only be resolved by assuming that the low temperature paramagnetism is parasitic and stems from paramagnetic impurities.

Conversely, if the intrinsic magnetic behavior of MnB$_4$ would be paramagnetic (as stated by Gou et al. \cite{gou2014}), in order of being covered by the diamagnetic signal of the sample holder/Apiezon it would require an extremely weak paramagnetic behavior ($\chi$ being of the order 10$^{-5}$\,emu/mole). Such a behavior is still inconsistent with Gou et al. \cite{gou2014}. While we consider this scenario unlikely, one might speculate about Pauli paramagnetism with a very small density of states at the Fermi level producing such a magnetic signature, {\it i.e.}, a pseudo-gap system. Again, however, this would not explain the Curie like paramagnetic increase of the susceptibility at low temperatures, which would have to be attributed to paramagnetic impurities.

In consequence, regarding the magnetic behavior of MnB$_4$, the most likely scenario is the following: The high temperature behavior likely reflects the intrinsic behavior of (pseudo-gapped) MnB$_4$. At low temperatures, residual magnetic (Mn) impurities produce a weakly paramagnetic/ferromagnetic signal. If we assume a typical moment of a few $\mu_B$ per Mn atom for these impurities, less than 0.1\% Mn impurities would be sufficient to explain the observed behavior. From our data we cannot unambiguously assess if we are dealing with dilute impurities dissolved in the crystallographic structure of MnB$_4$, or if they are assembled as clusters in grain boundaries etc.. Only the observation that the transition to a paramagnetic susceptibility occurs at a comparatively large temperature close to 100\,K seems to be more in line with larger clusters rather than dissolved impurities in a matrix.

\section{CONCLUSION}
\label{conclusion}

Summarizing our findings on the electronic and magnetic ground state of MnB$_4$, we have established that the electronic transport properties can be attributed to a pseudo-gap in the density of states at the Fermi-level. This observation is fully consistent with the theoretical analysis presented in Knappschneider et al. \cite{knappschneider1}. Further, we have demonstrated that at room temperature the magnetic susceptibility is either diamagnetic or very weakly paramagnetic, again fully consistent with the concept of a pseudo-gapped material. With all likelyhood, a paramagnetic low-temperature signal just reflects a response from impurities with an atomic density of a fraction of a percent.  

Notably, we find no evidence for a correlated metallic state as proposed by Gou et al. \cite{gou2014}. Given that in both studies \cite{knappschneider1,gou2014} the same crystallographic structure has been reported, the scenario of a structural instability producing different types of MnB$_4$ at low temperatures, as implicitly suggested based on band structure calculations \cite{liang,LiangWuWang}, appears not to explain the observed behavior. Rather, based on our experimental findings, residual impurities in polycrystalline material seem sufficient to account for the observed behavior. Or in other words, from our study the real material MnB$_4$ emerges to be intrinsically a ''dirty'' pseudo-gap system.

Still, the sensibility of the different band structure calculations on the chosen crystallographic structure and magnetic polarization is quite remarkable \cite{gou2012,litterscheid,yang,knappschneider1,gou2014,niu,geest,liang,knappschneider2,LiangWuWang}. Notably, the structural instability proposed to exist in MnB$_4$ \cite{liang,LiangWuWang} deserves further investigation. From these calculations, it appears that the structural stability at higher temperatures of MnB$_4$ is closely linked to its electronic properties, {\it viz.}, that electron-phonon coupling requires close attention in modelling this material. 

\section*{Acknowledgements} 

We gratefully acknowledge support by the Braunschweig International Graduate School of Metrology B-IGSM and the DFG Research Training Group GrK1952/1 ''Metrology for Complex Nanosystems''.




\begin{thebibliography}{00}
\bibitem{crystal} In Fig. 1 the crystals are shown prior to preparation for resistivity measurements. After embedding the samples the effective surface contact areas of the MnB$_4$ crystals are $\sim 150$\,$\mu$m$^2$ and larger.
\end{thebibliography}


\end{document}